\documentstyle[12pt]{article}
\newcommand{\fcaption}[1]{
        \addtocounter{figure}{1}
         {{\tenrm Fig.~\thefigure . #1} }\hfil\break }

\newcommand{\tcaption}[1]{                      
        \addtocounter{table}{1}
         {{\tenrm\offinterlineskip Table~\thetable . #1} }\hfil\break }

\newcommand{\nonumsection}[1] {\vspace{12pt}\noindent{\bf #1}
        \par\vspace{5pt}}

\newcommand{\be}{\begin{eqnarray}}
\newcommand{\ee}{\end{eqnarray}}
\newcommand{\dslash}{\partial \hskip -0.5em /}
\newcommand{\Dslash}{D \hskip -0.7em /}
\newcommand{\Gslash}{\Gamma \hskip -0.7em /}
\newcommand{\tr}{{\rm tr}}

\newcommand{\A}{{\cal A}}

\newcommand{\textlineskip}{\baselineskip=14pt}

 1
 1
 1

\begin{document}
\input psfig

{\thispagestyle{empty}}
\setcounter{page}{1}

\rightline{UNITU-THEP-6/1994}
\rightline{April 1994}
\rightline{hep-ph/9404222}

\vspace{1cm}
\centerline{\Large \bf Topologically non--trivial chiral
transformations}
\vspace{0.5cm}
\centerline{\Large \bf and their representations in a finite model
space$^\dagger $}
\vspace{0.5cm}
\centerline{R.\ Alkofer, H.\ Reinhardt, J.\ Schlienz and
H.\ Weigel$^\ddagger $}
\vspace{0.2cm}
\centerline{Institute for Theoretical Physics}
\vspace{0.2cm}
\centerline{T\"ubingen University}
\vspace{0.2cm}
\centerline{Auf der Morgenstelle 14}
\vspace{0.2cm}
\centerline{D-72076 T\"ubingen, FR Germany}

\vspace{2cm}
\centerline{\bf Abstract}
\vspace{0.5cm}
\baselineskip14pt
\noindent
The role of chiral transformations in effective theories
modeling Quantum Chromo Dynamics is reviewed. In the context
of the Nambu--Jona--Lasinio  model the hidden gauge and massive
Yang--Mills approaches to vector mesons are demonstrated to be
linked by a special chiral transformation which removes the chiral
field from the scalar--pseudoscalar sector. The role of this
transformation in the presence of a topologically non--trivial
chiral field is illuminated. The fermion determinant for such a field
configuration is evaluated by summing the discretized eigenvalues of
the Dirac Hamiltonian. This discretization is accomplished by
demanding certain boundary conditions on the quark fields leaving a
finite model space. The properties of two sets of boundary conditions
are compared. When the topologically non-trivial chiral transformation
is applied to the meson fields the associated transformation of the
boundary conditions is shown to be indispensable. A constructive
procedure for transforming the boundary conditions is developed.

\vfill
\noindent
$^\dagger $
{\footnotesize{Supported by the Deutsche Forschungsgemeinschaft (DFG)
under contract  Re 856/2-1.}}
\newline
$^\ddagger $
{\footnotesize{Supported by a Habilitanden--scholarship of the DFG}}
\eject

\normalsize\textlineskip
\section{Introduction}

Since a solution to Quantum Chromo Dynamics (QCD) is not yet
available one has to recede on models in order to explore processes
described by the strong interaction. These models are usually
constructed under the requirement that the symmetries of the
underlying theory, {\it i.e.} QCD are maintained. In this context
chiral symmetry and its spontaneous breaking play a key role.
In this article we will explore a special chiral transformation
when topologically non--trivial meson field configurations like
solitons are involved. To begin with, let us briefly review the
relevance of chiral symmetry on the one side and solitonic field
configurations on the other in the context of strong interactions.

QCD can be extended from $SU(3)$ to $SU(N_C)$ where $N_C$
denotes the number of color degrees of freedom. It was observed
by `t Hooft\cite{tho74} that in the limit $N_C\rightarrow\infty$
QCD is equivalent to an effective theory of weakly interacting mesons.
Subsequently Witten\cite{wi79} conjectured that baryons emerge as
solitons of the meson fields within this effective theory. Stimulated
by Witten's conjecture much interest has been devoted to the
description of baryons as chiral solitons during the past
decade\cite{ad83,ho93}. In the soliton description of baryons the
chiral field and in particular its topological structure play a key
role. The topological character of the chiral field especially endows
the soliton with baryonic properties like baryon charge and
spin\cite{wi83}. This comes about via the chiral anomaly\cite{ba69}
which is a unique feature of all quantum field theories where fermions
live in a gauge group and couple to a chiral field. The fermionic part
of such a theory has the generic structure
\be
Z_F[\Phi]
&=& \int D \Psi D \bar{\Psi}\ \mbox{exp}\left[
i\int\ d^4x\ \bar{\Psi}\left(i \Dslash\ -\hat m_0\right)
\Psi\right]  \nonumber\\
&=& \mbox{Det}\, \left(i\Dslash\ -\hat m_0\right)
\label{zferm}
\ee
where $\hat m_0$ denotes the current quark mass matrix which will
be ignored in the ongoing discussions. Furthermore
\be
i\Dslash
= i \dslash- \Phi
= i(\dslash + \Gslash ) - MP_R - M^\dagger P_L
\label{Dslash}
\ee
represents the Dirac--operator of the fermions in the external
Bose--field $\Phi$. $\Phi$ in general contains vector $V_\mu$,
axial vector $A_\mu$ fields\footnote{In our notation $V_\mu$ and
$A_\mu$ are anti--hermitian.} as well as scalar $S$ and
pseudo-scalar fields $P$
\be
\Gamma_\mu = V_\mu + A_\mu\gamma_\mu,
\qquad M=S+iP=\xi_L^\dagger \Sigma \xi_R.
\label{vecfield}
\ee
Here $P_{R/L} =  \frac{1}{2}(1\pm\gamma_5)$ are the chiral
projectors. Accordingly one defines left(L)-- and
right(R)--handed quark fields: $\Psi_{L,R}=P_{L,R}\Psi$. The
chiral field $U$ is defined via the polar decomposition
of the meson fields
\be
U = \xi_L^\dagger \xi_R = {\rm exp}(i\Theta)
\label{chifield}
\ee

The chiral anomaly arises because there is no regularization
scheme which simultaneously preserves local vector and axial
vector (chiral) symmetries. In renormalizable theories the
chiral anomaly can be calculated in closed form and is given
by the Wess-Zumino action\cite{we79,wi83,kay84}
\be
{\cal A} = S_{WZ} = \frac{iN_C}{240\pi^2}
\int_{M^5}(UdU^\dagger)^5
\label{wzw}
\ee
where the notation of alternating differential forms has been used.
$M^5$ denotes a five dimensional manifold whose boundary is
Minkowski space. Obviously the chiral anomaly is tightly related
to the chiral field since the Wess-Zumino action vanishes when
the chiral field disappears $(U = 1)$. The chiral anomaly is,
however, not merely a technical artifact but has well established
physical consequences.  In the meson sector it gives rise to the
so--called ``anomalous decay processes'' like {\it e.g.}
$\pi \rightarrow 2\gamma$ and $\omega \rightarrow 3\pi$. In the
soliton sector the chiral anomaly requires for $N_C = 3$ the soliton
to be quantized as a fermion and endows the soliton with half integer
spin and integer baryon number\footnote{For this proof it is
mandatory to consider flavor SU(3).}\cite{wi83,ma84}.

For many purposes it is convenient to perform a chiral rotation
of the fermions\cite{eb86,re89a}
\be
\tilde{\Psi} = \Omega \Psi\quad {\rm with} \qquad
\Omega = P_L \xi_L + P_R\xi_R,
\quad {\it i.e.}\quad \tilde{\Psi}_{L,R}=\xi_{L,R}\Psi_{L,R}.
\label{psitilde}
\ee
This transformation defines a chirally rotated Dirac-operator
\be
\Psi i \Dslash\,\Psi = \bar{\tilde{\Psi}} i \tilde{\Dslash} \, \tilde{\Psi}
\label{rot1}
\ee
which acquires the form
\be
i\tilde{\Dslash} = \Omega^\dagger i \Dslash \, \Omega^\dagger =
i\gamma_\mu\left(\partial^\mu+\tilde{V}^\mu
+\tilde{A}^\mu\gamma_5\right)-\Sigma.
\label{rot2}
\ee
The chiral rotation has removed the chiral field from the scalar
pseudo-scalar sector of the rotated Dirac operator $i\tilde{\Dslash}$.
As a consequence the vector and axial vector fields become now
chirally rotated
\be
\tilde{V}_\mu + \tilde{A}_\mu
&=& \xi_R(\partial_\mu + V_\mu + A_\mu) \xi_R^\dagger, \nonumber\\
\tilde{V}_\mu - \tilde{A}_\mu
&=& \xi_L(\partial_\mu + V_\mu - A_\mu) \xi_L^\dagger.
\label{rotvec}
\ee
Even in the absence of vector and axial vector
fields in the original Dirac operator $(V_\mu = A_\mu = 0)$
the chiral rotation induces vector and axial vector fields
\be
\tilde{V}_\mu (V_\mu = A_\mu = 0)=v_\mu
&=&\frac{1}{2}\left(\xi_R \partial_ \mu \xi_R^\dagger +
\xi_L \partial_\mu \xi_L^\dagger\right),
    \nonumber\\
\tilde{A}_\mu (V_\mu = A_\mu = 0)=a_\mu
&=& \frac{1}{2}\left(\xi_R \partial_\mu \xi_R^\dagger -
\xi_L\partial_\mu \xi_L^\dagger\right).
\label{indvec}
\ee

For the soliton description of baryons the chiral field is usually
assumed to be of the hedgehog type
\be
U = {\rm exp}\left(
i\Theta(r){\mbox{\boldmath $\tau$}}
\cdot{\hat{\mbox{\boldmath $r$}}}\right)
\label{hedgehog}
\ee
The non--trivial topological structure of this configuration is then
exhibited by the boundary conditions $\Theta(0)=-n\pi$ and
$\Theta(\infty)=0$. The chiral field thus represents a mapping
from the compactified coordinate space (all points at spatial infinity
are identified) to SU(2) flavor space, {\it i.e.} $S^3\rightarrow S^3$.
The associated homotopy group, $\Pi_3(S^3)$, is isomorphic to $Z$, the
group of integer numbers. The isomorphism is given by the winding number
$\left(\Theta(0)-\Theta(\infty)\right)/\pi=-n$. Assuming the unitary
gauge $(\xi_L^\dagger=\xi_R)$  the induced vector field
is of the Wu--Yang form\cite{re89a}
\be
v_0 = 0, \quad
v_i = i v_i^a \frac{\tau^a}{2}, \quad
v_i^a = \epsilon^{ika} \hat{r}_k \frac{G(r)}{r}
\label{wuyang}
\ee
with the profile function $G(r)$ given by the chiral angle $\Theta(r)$
\be
G(r) = -2 \sin^2 \frac{\Theta(r)}{2}.
\label{gtheta}
\ee
For odd $n$ the topological non--trivial character of the chiral
rotation is also reflected by a non--vanishing value of the
induced vector field at $r=0$.
The induced axial vector field $a_i = i a_i^a \tau^a/2$
becomes
\be
a_i^a =\hat{r}_i \hat{r}_a
\left(\Theta^\prime(r) - \frac{\sin\Theta(r)}{r}\right)
+ \delta_{ia} \frac{\sin\Theta(r)}{r}
\label{atheta}
\ee
where the prime indicates the derivative with respect to the argument.

The use of the chirally rotated fermions is advantageous since the
rotated Dirac-operator does no longer contain a chiral field and
its determinant is hence anomaly free. The chiral anomaly,
however, has not been lost by the chiral rotation but is hidden
in the integration measure over the fermion fields. In fact, as
observed by Fujikawa\cite{fu80}, a chiral rotation of the fermion
fields gives rise to a non--trivial Jacobian of the integration
measure
\be
D\Psi D\bar{\Psi} = J(U) D \tilde{\Psi} D \bar{\tilde{\Psi}}
\label{measure}
\ee
which is precisely given by the chiral anomaly
\be
J(U) = \mbox{exp}\,\left(i\cal{A}\right)
\label{measano}
\ee
Thus we have the relation
\be
- i {\rm Tr} \log i \Dslash = - i {\rm Tr} \log i\tilde{\Dslash} + \cal{A}
\label{actiontrans}
\ee
In many cases it is convenient to work with the chirally rotated
fermion fields because of the absence of the anomaly from the
fermion determinant.

As already mentioned above the chiral anomaly or equivalently the
non--trivial Jacobian in the fermionic integration measure arise
due to the need for regularization, which introduces a finite cut--off.
In the regularized theory the chiral anomaly can be evaluated in a
gradient expansion. In leading order the anomaly is then given by
the Wess-Zumino action (\ref{wzw}). There are, however, higher order
terms which are suppressed by inverse powers of the cut--off. In
renormalizable theories where the cut--off goes to infinity, these
higher order terms disappear and the chiral anomaly is known in closed
form. In non--renormalizable effective theories, however, the cut--off
of the regularization scheme has to be kept finite and acquires a
physical meaning, indicating the range of validity of the effective
theory. In this case the higher order terms of the gradient expansion
do not disappear but contribute to the anomaly which is then no longer
available in closed form.

Furthermore, when the soliton sector of such effective mesonic
theories is studied it is not sufficient to only consider the leading
and sub--leading contributions from the gradient expansion and one
has to perform a full non--perturbative evaluation of the fermion
determinant\cite{re89}. The non--perturbative calculations have to
be performed numerical\-ly\cite{re88}, with the continuous space being
discretized. Also in the non--perturbative studies of the soliton
sector of the effective theory the use of the chiral rotation is in
many cases advantageous\cite{re89a}. As noticed above for the soliton
description of baryons the topological nature of the chiral field is
crucial. Actually, topology is a property of continuous spaces
(manifolds) and it is {\it a priori} not clear whether the chiral
rotation with a topologically non--trivial chiral field can be
represented in a finite dimensional and discretized model space used
in the numerical calculations. In this respect let us recall that a
single point defect in a manifold changes its topological properties
already drastically.

The present paper is devoted to a study of the chiral rotation
in non--perturbative soliton calculations where the fermion
determinant has to be numerically evaluated in the background of
a topologically non--trivial chiral field. For definiteness
we shall use the Nambu-Jona-Lasinio (NJL)\cite{na61} model as
a microscopic fermion theory which shares all the relevant
properties of chiral dynamics with QCD. Its bosonized version
gives a quite satisfactory description of mesons\cite{eb86} and also
of baryons when the soliton picture is assumed\cite{al93}. The
organization of the paper is as follows: After these introductory
remarks we review the importance of the local chiral rotation
(\ref{psitilde}) for the extraction of meson properties from the
NJL action. In section 3 we discuss the soliton solution to the NJL
model of pseudoscalar fields with special emphasis on the choice of
boundary conditions in the finite model space. Section 4 finally is
devoted to the study of the local chiral rotation for topologically
non--trivial field configurations and its influence on the soliton
solution. Concluding remarks are given in section 5.

\section{Properties of local chiral transformations}

As pointed out in the introduction the chirally rotated formulation
of the NJL model is suited to investigate properties of (axial--)
vector mesons. In the present section we will therefore briefly
review the results and illuminate the connection with the
hidden gauge symmetry (HGS) and massive Yang--Mills (MYM) approaches
for the description of (axial--) vector mesons. Most of the
results reported in this section are taken from earlier
works\cite{eb86,wa88,re89a,wa89}. Nevertheless we repeat these
results here in order to put our work into perspective and have the
paper self--contained.

In the chirally rotated formulation the bosonized version of the
NJL model action reads
\be
{\cal A}_{\rm NJL}=- i {\rm Tr} \log i\tilde{\Dslash} + {\cal A}
-\frac{1}{4G_1}{\rm tr} \left(\Sigma^2 - m^2 \right)
-\frac{1}{4G_2}{\rm tr} \left[\left(\tilde V_\mu-v_\mu\right)^2
+\left(\tilde A_\mu-a_\mu\right)^2\right]
\label{njlrot}
\ee
with the constituent quark mass $m$ being the vacuum expectation value
of the scalar field $\Sigma$ {\it i.e.} $\langle\Sigma\rangle=m$. Again
we have discarded terms proportional to the current quark mass matrix.
The chirally transformed (axial--) vector fields are defined in eqns
(\ref{rotvec}) and (\ref{indvec}). Next we have to face the fact
that the functional trace of the logarithm in (\ref{njlrot}) is
ultra--violet divergent and thus needs regularization. This is
achieved by first continuing to Euclidean space $(x_0=-ix_4)$ and then
representing the real part of the Euclidean action by a parameter
integral
\be
\frac{1}{2}{\rm Tr} \log \left( \tilde{\Dslash}_E^\dagger
\tilde{\Dslash}_E\right) \longrightarrow
-\frac{1}{2}\int_{1/\Lambda^2}^\infty \frac{ds}{s}
{\rm exp}\left(-s \tilde{\Dslash}_E^\dagger
\tilde{\Dslash}_E\right)
\label{regreal}
\ee
which introduces the cut--off $\Lambda$. This substitution is an
identity up to an irrelevant constant for $\Lambda\rightarrow\infty$.
The Euclidean Dirac operator $\tilde{\Dslash}_E$ is obtained by
analytical continuation of $\tilde{\Dslash}$ to Euclidean space.
The prescription (\ref{regreal}) is known as the proper--time
regularization\cite{sch51}. For the purpose of the present paper
it is sufficient to only consider the normal parts of the action.
We may therefore neglect the imaginary part as well as the anomaly
${\cal A}$. Thus the actual starting point of our considerations is
represented by
\be
{\cal A}_{\rm NJL}&=&-\frac{1}{2}\int_{1/\Lambda^2}^\infty \frac{ds}{s}
{\rm Tr}\ {\rm exp}\left(-s \tilde{\Dslash}_E^\dagger
\tilde{\Dslash}_E\right)
-\frac{1}{4G_1}{\rm tr} \left(\Sigma^2 - m^2 \right)
\nonumber \\ &&\hspace{3cm}
-\frac{1}{4G_2}{\rm tr} \left[\left(\tilde V_\mu-v_\mu\right)^2
+\left(\tilde A_\mu-a_\mu\right)^2\right].
\label{stpoint}
\ee
In order to extract information about the properties of the (axial--)
vector mesons commonly a (covariant) derivative expansion of the
fermion determinant is performed. We choose to consider a covariant
derivative expansion since, in contrast to an on--shell determination
of the parameters\cite{ja92}, it preserves gauge invariance and does
not lead to artificial mass terms. Furthermore, this procedure leaves
the extraction of the axial--vector meson mass unique. Continuing back
to Minkowski space and substituting the scalar field $\Sigma$ by its
vacuum expectation value yields the leading terms\cite{wa88,re89a}
\be
{\cal L}_{\rm NJL}=\frac{1}{2g_V^2}{\rm tr}\
\left(\tilde V_{\mu\nu}^2+\tilde A_{\mu\nu}^2\right)
-\frac{6m^2}{g_V^2}{\rm tr}\ \tilde A_\mu^2
-\frac{1}{4G_2}{\rm tr} \left[\left(\tilde V_\mu-v_\mu\right)^2
+\left(\tilde A_\mu-a_\mu\right)^2\right] + \ldots\ .
\label{expand}
\ee
Here $\tilde V_{\mu\nu}$ and $\tilde A_{\mu\nu}$ denote the field
strength tensors
\be
\tilde V_{\mu \nu }&=& \partial_\mu\tilde V_\nu-\partial_\nu \tilde
V_\mu+[\tilde V_\mu, \tilde V_\nu]+[\tilde A_\mu, \tilde A_\nu],
\nonumber \\*
\tilde A_{\mu \nu} &=& \partial_\mu \tilde A_\nu - \partial_\nu
\tilde A_\mu+[\tilde V_\mu, \tilde A_\nu]+[\tilde A_\mu, \tilde V_\nu]
\label{ftensor}
\ee
of vector and axial--vector fields, respectively. Obviously in our
convention these fields contain the coupling constant $g_V$ which
in the proper--time regularization is given by
\be
g_V=4\pi\left[\frac{2N_C}{3}
\Gamma\left(0,\left(\frac{m}{\Lambda}\right)^2\right)\right]
^{-\frac{1}{2}}.
\label{copconst}
\ee
For the description of the pion fields we adopt the unitary
gauge for the chiral fields: $\xi=\xi^\dagger_L=\xi_R$. The
pions come into the game by the non--linear realization
$\xi={\rm exp}\left(i\mbox{\boldmath $\tau$}\cdot\mbox{\boldmath
$\pi$}/f\right)$. Then the last term in eqn (\ref{expand}) contains
the axial--vector pion mixing which is eliminated by a corresponding
shift in the axial field: $\tilde A_\mu\rightarrow\tilde A^\prime_\mu
=\tilde A_\mu+(ig_V^2m^2/12fG_2)\partial_\mu
\mbox{\boldmath $\tau$}\cdot\mbox{\boldmath $\pi$}$. This shift
obviously provides an additional kinetic term for the pions and thus
effects the pion decay constant
\be
f_\pi^2=\frac{M_A^2-M_V^2}{4M_A^2G_2}
\label{fpi}
\ee
with the (axial--) vector meson masses
\be
M_V^2=\frac{g_V^2}{4G_2}\qquad {\rm and}\qquad
M_A^2=M_V^2+6m^2.
\label{vecmass}
\ee
This brief summary of known results has demonstrated the usefulness
of the chiral rotation (\ref{psitilde}) especially in the context
of the derivative expansion since it eliminates the derivative of
the chiral field from the fermion determinant (\ref{rot2}).

The Lagrangian of the hidden gauge approach can be obtained from
(\ref{expand}) by the following approximation. One neglects the
kinetic parts for the axial--vector field $\tilde A_\mu^\prime$ which
leaves this field only as an auxiliary field. This allows to employ
the corresponding equation of motion to eliminate $\tilde A_\mu^\prime$
resulting in
\be
{\cal L}\sim\frac{1}{2g_V^2}{\rm tr}\ \tilde V_{\mu\nu}^2
-a f_\pi^2\ {\rm tr}\ \left(\tilde V_\mu -v_\mu\right)^2
-\frac{1}{4}f_\pi^2\ {\rm tr}\ a_\mu^2.
\label{bando}
\ee
In the work of Bando et al.\cite{ba87} $a$ was left as an undetermined
parameter. Here it is fixed in terms of physical quantities
\be
a=\frac{M_A^2}{M_A^2-M_V^2}.
\label{apara}
\ee
Assuming the constituent quark mass $m=M_V/\sqrt{6}$ not only
yields the Weinberg relation $M_A=\sqrt{2}M_V$\cite{we67} but
also the KSRF relation $a=2$\cite{ka66}.

Alternatively one might apply the same manipulations to the
formulation in terms of the unrotated fields (\ref{vecfield}).
Then the chiral field still appears in the fermion determinant
and one has to deal with the covariant derivative
\be
{\cal D}_\mu U = \partial_\mu U +\left[V_\mu,U\right]
-\left\{A_\mu,U\right\}.
\label{covder}
\ee
The leading terms in the Lagrangian can readily be obtained\cite{eb86}
\be
{\cal L}\sim \frac{3m^2}{2g_V^2}{\rm tr}\
\left({\cal D}_\mu U {\cal D}^\mu U^\dagger\right)
+\frac{1}{2g_V^2}{\rm tr}\
\left(V_{\mu\nu}^2+A_{\mu\nu}^2\right)
-\frac{1}{4G_2}{\rm tr} \left(V_\mu^2+A_\mu^2\right)
\label{expunrot}
\ee
which exactly represent the massive Yang--Mills
Lagrangian\cite{kay84,go84}. Transforming the (axial--) vector
fields according to (\ref{rotvec}) and noting that\cite{wa88}
\be
{\rm tr}\ \tilde A_\mu^2 = \frac{-1}{4}{\rm tr}\
\left({\cal D}_\mu U {\cal D}^\mu U^\dagger\right)
\label{aeqdu}
\ee
one immediately observes that (\ref{expunrot}) and (\ref{expand})
describe the same physics. In the language of the NJL model the
identity of the HGS and MYM approaches stems form the invariance
of the module of the fermion determinant under the special chiral
rotation (\ref{psitilde}).

Let us next explore the behavior of the fields under flavor rotations
$g_L,\ g_R$. These are defined for the unrotated left-- and
right--handed quark fields
\be
\Psi_L\rightarrow g_L\Psi_L\qquad {\rm and} \qquad
\Psi_R\rightarrow g_R\Psi_R.
\label{qtransf}
\ee
The term which describes the coupling of the quarks to
the scalar and pseudoscalar mesons is left invariant by demanding
\be
\xi_L^\dagger\Sigma\xi_R\rightarrow
g_L\xi_L^\dagger\Sigma\xi_R g_R^\dagger
\label{mtransf}
\ee
which introduces the hidden gauge transformation $h$\cite{eb86}
\be
\xi_L\rightarrow h^\dagger \xi_L g_L^\dagger
\quad , \qquad
\xi_R\rightarrow h^\dagger \xi_R g_R^\dagger
\qquad {\rm and}\qquad
\Sigma\rightarrow h^\dagger\Sigma h.
\label{xitransf}
\ee
Obviously the scalar fields transform homogeneously under the
hidden gauge transformation. In this context it is important to
note that $h$ may not be chosen independently but rather
depends on the gauge adopted for the chiral fields. Consider
{\it e.g.} the unitary gauge $\xi_L^\dagger=\xi_R=\xi$.
This requires the transformation property
\be
\xi\rightarrow g_L\xi h = h^\dagger \xi g_R^\dagger.
\label{ugaugetrans}
\ee
For vector type transformations $g_L=g_R=g_V$ this equation is
obviously solved by $h=g_V^\dagger$. Contrary, for axial type
transformations $g_L=g_R^\dagger=g_A$ $h$ is obtained as the
solution to $g_A\xi h = h^\dagger \xi g_A$ which depends on
the field configuration $\xi$. Thus even for global flavor
transformations $g_{A,V}$ the hidden gauge transformation $h$ may be
coordinate--dependent for coordinate dependent
$\xi(x)$\cite{ca69,kay84a}. The unrotated (axial--) vector fields
transform inhomogeneously under the flavor rotations
\be
V_\mu+A_\mu\rightarrow
g_R\left(\partial_\mu + V_\mu+A_\mu\right)g_R^\dagger
\qquad {\rm and}\qquad
V_\mu-A_\mu\rightarrow
g_L\left(\partial_\mu + V_\mu-A_\mu\right)g_L^\dagger.
\label{vtrans}
\ee
It is then straightforward to verify that the flavor transformation
of the rotated fields only involves the hidden symmetry transformation
$h$\cite{eb86}
\be
\tilde \Psi_{L,R}\rightarrow h^\dagger \tilde \Psi_{L,R}
\quad , \qquad
\tilde V_\mu\rightarrow
h^\dagger\left(\partial_\mu +\tilde V_\mu\right)h
\qquad {\rm and}\qquad
\tilde A_\mu\rightarrow h^\dagger \tilde A_\mu h.
\label{vrottrans}
\ee
The fact that $\tilde A_\mu$ transforms homogeneously has the
important consequence that one can build models by setting
$\tilde A_\mu\equiv0$ without breaking chiral symmetry. These
models then do no longer contain axial--vector degrees of
freedom. Such models have frequently been utilized in the
investigation of $\rho$ and $\omega$ meson physics\cite{kay84a,sch93}.
They have also been successfully applied to describe
baryons as solitons\cite{ja88,me88,sch89}.

\section{The NJL soliton}

In the baryon number one sector the NJL model has the celebrated
feature to possess localized static solutions with finite energy,
{\it i.e.} solitons\cite{re89,re88}. Here we wish to briefly
review this solution and discuss different boundary conditions for
the quark wave--functions.

For static field configurations it is convenient to introduce
a Dirac Hamiltonian
\be
{\cal H}= \mbox {\boldmath $\alpha \cdot p $} +
\beta\left(P_R\xi\langle \Sigma\rangle\xi
+P_L\xi^{\dag}\langle \Sigma\rangle\xi^{\dag}\right)
\label{stham}
\ee
where we have assumed the unitary gauge ({\it i.e.}
$\xi_L^{\dag}=\xi_R=\xi$). This Hamiltonian enters the Euclidean
Dirac operator via
\be
i\beta\Dslash_E=-\partial_\tau-{\cal H}.
\label{eucdirac}
\ee
For static mesonic background fields the fermion determinant can
conveniently be expressed in terms of the eigenvalues $\epsilon_\mu$
of the Dirac Hamiltonian
\be
{\cal H}\Psi_\mu=\epsilon_\mu\Psi_\mu.
\label{direig}
\ee
These eigenvalues obviously are functionals of the mesonic background
fields. Depending of the specific boundary condition (which fixes
the quantum reference state) to the fermion fields in the functional
integral (\ref{zferm}), the fermion determinant (\ref{zferm})
contains in general besides a vacuum part ${\cal A}^0$ also
a valence quark part ${\cal A}^{\rm val}$\cite{re89}
\be
{\cal A}={\cal A}^0+{\cal A}^{\rm val}.
\label{actionsum}
\ee
The valence quark part arising from the explicit occupation of
the valence quark levels is given by
\be
{\cal A}^{\rm val}=-E^{\rm val}[\xi] T\ , \quad
E^{\rm val}[\xi]=N_C\sum_\mu \eta_\mu |\epsilon_\mu|.
\label{eval}
\ee
Here $\eta_\mu=0,1$ denote the occupation numbers of the
valence (anti-) quark states. The vacuum part is conveniently evaluated
for infinite Euclidean times $(T\rightarrow\infty)$ which fixes
the vacuum state as the quantum reference state (no valence quark
orbit occupied). For the present considerations it will be sufficient
to evaluate the real vacuum part
\be
{\cal A}^0_R=\frac{1}{2} {\rm Tr}\ {\rm log} \Dslash^\dagger_E
\Dslash_E.
\label{areal}
\ee
Since for static configurations one has $[\partial_\tau,{\cal H}]$=0
and thus $\Dslash_E ^{\dag}\Dslash_E = -\partial_\tau^2 +{\cal H}^2$.
Then it is straightforward to evaluate the real part of the
fermion determinant in proper--time regularization\footnote{The
imaginary part does not contribute for the field configurations
under consideration.}\cite{re89}
\be
\A^0_R &=& -\frac{1}{2}\int_{1/\Lambda ^2}^\infty \frac {ds}{s}
{\rm Tr}\ {\rm exp}
\left( -s \Dslash_E ^{\dag}\Dslash_E \right)
\nonumber \\
&=& -T \frac{N_C}{2}\int_{-\infty}^\infty \frac{dz}{2\pi}\sum_\mu
\int_{1/\Lambda ^2}^\infty \frac {ds}{s}
{\rm exp}\left(-s\left[z^2+\epsilon_\mu^2\right]\right).
\label{star}
\ee
The temporal part of the trace has become the $z$ integration. As
this integral is Gaussian it can readily be carried out yielding
\be
\A^0_R=-T \frac{N_C}{2}\int_{1/\Lambda ^2}^\infty
\frac{ds}{\sqrt{4\pi s^3}}\sum_\mu
{\rm exp}\left(-s\epsilon_\mu^2\right).
\label{arsum}
\ee
This expression allows to read off the static energy functional $E[\xi]$
since $\A^0_R\rightarrow -T E^0[\xi]$ as $T\rightarrow\infty$
\be
E^0[\xi]=\frac{N_C}{2}\int_{1/\Lambda ^2}^\infty
\frac{ds}{\sqrt{4\pi s^3}}\sum_\mu
{\rm exp}\left(-s\epsilon_\mu^2\right).
\label{evac}
\ee
The occupation numbers $\eta_\mu$ of the valence quark orbits have to
adjusted such the baryon number
\be
B=\sum_\mu \left(\eta_\mu
-\frac{1}{2}{\rm sign}\left(\epsilon_\mu\right)\right)
\label{bno}
\ee
equals unity. The total energy functional is finally given by
\be
E[\xi]=E^{\rm val}[\xi]+E^{\rm vac}[\xi]-E^0[\xi=1]
\label{etot}
\ee
which is normalized to the energy of the vacuum configuration
$\xi=1$. In the chiral limit ($m_\pi=0$), which we have
adopted here, the meson part of the action does not contribute
to the soliton energy. The chiral soliton is the $\xi$ configuration
which minimizes $E[\xi]$ and the minimal $E[\xi]$ is then identified
as the soliton mass.

To be specific we employ the hedgehog {\it ansatz} for the chiral
field
\be
\xi({\bf r})={\rm exp}\left(\frac{i}{2}
{\mbox{\boldmath $\tau$}} \cdot{\bf \hat r}\ \Theta(r)\right)
\label{chsol}
\ee
while the scalar fields are constrained to the chiral circle,
{\it i.e.} $\langle \Sigma\rangle =m$. Substituting this
{\it ansatz} into the Dirac Hamiltonian (\ref{stham}) yields
\be
{\cal H}&=&
{\mbox {\boldmath $\alpha \cdot p $}} +
\beta m \left({\rm cos}\Theta(r) + i\gamma_5{\mbox{\boldmath $\tau$}}
\cdot{\bf \hat r}\ {\rm sin}\Theta(r)\right).
\label{h0}
\ee

The stationary condition $\delta E[\xi]/\delta\xi=0$ is made
explicit by functionally differentiating the energy--eigenvalues
$\epsilon_\mu$ with respect to $\Theta$
\be
\frac{\delta\epsilon_\mu}{\delta\Theta(r)}=
m \int d\Omega\  \Psi_\mu^\dagger({\bf r})\beta
\left(-{\rm sin}\Theta(r)+i\gamma_5{\mbox{\boldmath $\tau$}}
\cdot{\bf \hat r}\ {\rm cos}\Theta(r)\right)\Psi_\mu({\bf r}).
\label{diffeps}
\ee
This leads to the equation of motion\cite{re88}
\be
{\rm cos}\Theta(r)\  {\rm tr}\int d\Omega\  \rho_S({\bf r},{\bf r})
i\gamma_5{\mbox{\boldmath $\tau$} \cdot{\bf \hat r}}\ =
{\rm sin}\Theta(r)\  {\rm tr}\int d\Omega\  \rho_S({\bf r},{\bf r})
\label{eqm}
\ee
where the traces are over flavor and Dirac indices only. According to
the sum (\ref{etot}) the scalar quark density matrix
$\rho_S({\bf x},{\bf y})=\langle q({\bf x})\bar q({\bf y})\rangle$
is decomposed into valence quark and Dirac sea parts:
\be
\rho_S(\mbox{\boldmath $x$}, \mbox{\boldmath $y$}) & = &
\rho_S^{\rm val}(\mbox{\boldmath $x$},\mbox{\boldmath $y$})
+ \rho_S^{\rm vac}(\mbox{\boldmath $x$},\mbox{\boldmath $y$})
\nonumber \\*
\rho_S^{\rm val}(\mbox{\boldmath $x$},\mbox{\boldmath $y$}) & = &
\sum _\mu
\Psi_\mu(\mbox{\boldmath $x$})\eta_\mu
\bar \Psi_\mu(\mbox{\boldmath $y$}) {\rm sign} (\epsilon_\mu)
\nonumber \\*
\rho_S^{\rm vac}(\mbox{\boldmath $x$},\mbox{\boldmath $y$}) & = &
\frac{-1}{2}\sum_\mu \Psi_\mu(\mbox{\boldmath $x$})
{\rm erfc}\left(\left|\frac{\epsilon_\mu}{\Lambda}\right|\right)
\bar \Psi_\mu(\mbox{\boldmath $y$}) {\rm sign} (\epsilon_\mu) .
\label{density}
\ee

Technically the discretized eigenvalues $\epsilon_\mu$ of the
Dirac Hamiltonian ${\cal H}$ (\ref{stham},\ref{direig}) are obtained by
restricting the space $R_3$ to a spherical cavity of radius $D$ and
demanding certain boundary conditions at $r=D$. Eventually the
continuum limit $D\rightarrow\infty$ has to be considered.
In order to discuss pertinent boundary conditions it is necessary
to describe the structure of the eigenstates of ${\cal H}$. Due to
the special form of the hedgehog {\it ansatz} the Dirac Hamiltonian
commutes with the grand spin operator
\be
{\bf G}={\bf J}+\frac{{\mbox{\boldmath $\tau$}}}{2}
=\mbox{\boldmath $l$}+
\frac{{\mbox{\boldmath $\sigma$}}}{2}+
\frac{{\mbox{\boldmath $\tau$}}}{2}
\label{gspin}
\ee
where ${\bf J}$ labels the total spin and $\mbox{\boldmath $l$}$
the orbital angular momentum. $\mbox{\boldmath $\tau$}/2$ and
$\mbox{\boldmath $\sigma$}/2$ denote the isospin and spin operators,
respectively. The eigenstates of ${\cal H}$ are then as well eigenstates
of ${\bf G}$. The latter are constructed by first coupling spin
and orbital angular momentum to the total spin which is subsequently
coupled with the isospin to the grand spin\cite{ka84}. The
resulting states are denoted by $|ljGM\rangle$ with $M$ being the
projection of ${\bf G}$. These states obey the selection rules
\be
j=\cases{G+1/2, & $l=\cases{G+1 &\cr G &}$ \cr
& \cr G-1/2, & $l=\cases{G &\cr G-1 &}$}.
\label{gstates}
\ee
The Dirac Hamiltonian furthermore commutes with the parity operator.
Thus the eigenstates of ${\cal H}$ with different parity and/or grand spin
decouple. The coordinate space representation of the eigenstates
$|\mu\rangle$ is finally given by
\be
\Psi_\mu^{(G,+)}=
\pmatrix{ig_\mu^{(G,+;1)}(r)|GG+\frac{1}{2}GM\rangle \cr
f_\mu^{(G,+;1)}(r) |G+1G+\frac{1}{2}GM\rangle \cr} +
\pmatrix{ig_\mu^{(G,+;2)}(r)|GG-\frac{1}{2}GM\rangle \cr
-f_\mu^{(G,+;2)}(r) |G-1G-\frac{1}{2}GM\rangle \cr}
\label{psipos} \\ \nonumber \\
\Psi_\mu^{(G,-)}=
\pmatrix{ig_\mu^{(G,-;1)}(r)|G+1G+\frac{1}{2}GM\rangle \cr
-f_\mu^{(G,-;1)}(r) |GG+\frac{1}{2}GM\rangle \cr} +
\pmatrix{ig_\mu^{(G,-;2)}(r)|G-1G-\frac{1}{2}GM\rangle \cr
f_\mu^{(G,-;2)}(r) |GG-\frac{1}{2}GM\rangle \cr}.
\label{psineg}
\ee
The second superscript labels the intrinsic parity $\Pi_{\rm intr}$
which enters the parity eigenvalue via $\Pi=(-1)^G\times\Pi_{\rm intr}$.
In the absence of the soliton ({\it i.e.} $\Theta=0$) the radial
functions $g_\mu^{(G,+;1)}(r),\ f_\mu^{(G,+;1)}(r),$ etc. are given by
spherical Bessel functions. {\it E.g.}
\be
g_\mu^{(G,+;1)}(r)=N_k\sqrt{1+m/E}\ j_G(kr),\quad
f_\mu^{(G,+;1)}(r)=N_k{\rm sign}(E)\sqrt{1-m/E}\ j_{G+1}(kr)
\label{freesol}
\ee
and all other radial functions vanishing represents a solution to
${\cal H}(\Theta=0)$ with the energy eigenvalues $E=\pm\sqrt{k^2+m^2}$
and parity $(-1)^G$. $N_k$ is a normalization constant.

Two distinct sets of boundary conditions have been considered in the
literature. Originally Kahana and Ripka\cite{ka84} proposed to
discretize the momenta by enforcing those components of the Dirac
spinors to vanish at the boundary which possess identical grand spin
and orbital angular momentum, {\it i.e.}
\be
g_\mu^{(G,+;1)}(D)=g_\mu^{(G,+;2)}(D)=
f_\mu^{(G,-;1)}(D)=f_\mu^{(G,-;2)}(D)=0.
\label{bc1}
\ee
This boundary condition has the advantage that for a given grand
spin channel $G$ only one set of basis momenta $\{k_{nG}\}$ is
involved. These $k_{nG}$ make the $G^{\rm th}$ Bessel function vanish
at the boundary ($j_G(k_{nG}D)=0$). However, this boundary condition
has (among others) the disadvantage that the matrix elements of
flavor generators, like $\tau_3$ are not diagonal in momentum
space. If the matrix elements of the flavor generators are not
diagonal in the momenta a finite moment of inertia will result even
in the absence of a chiral field\cite{re89}. Stated otherwise, in this
case the boundary conditions violate the flavor symmetry. This problem
can be cured\cite{we92} by changing the boundary conditions for the
states with $\Pi_{\rm intr}=-1$
\be
g_\mu^{(G,+;1)}(D)=g_\mu^{(G,+;2)}(D)=
g_\mu^{(G,-;1)}(D)=g_\mu^{(G,-;2)}(D)=0
\label{bc2}
\ee
{\it i.e.} the upper components of the Dirac spinors always vanish
at the boundary. The diagonalization of the Dirac Hamiltonian
(\ref{stham}) with the condition (\ref{bc2}) is technically less
feasible since it involves three sets of basis momenta
$\{k_{nG-1}\},\{k_{nG+1}\}$ and $\{k_{nG}\}$ for a given grand
spin channel. In table 1 we compare some properties of the
two boundary conditions (\ref{bc1}) and (\ref{bc2}) in the case
when no soliton is present. The first four quantities appearing
in that table show up in various equations of motion when {\it e.g.}
also (axial--) vector mesons are included\cite{sch93b}. In case such
a quantity is non--zero the vacuum gives a spurious contribution
to the associated equation of motion. For an iterative solution to
the equations of motion this spurious contribution has to be
subtracted. It should be noted, however, that the relations listed
in table 1 are all satisfied for both boundary conditions in the
continuum limit $D\rightarrow\infty$.

So far the discussion of the boundary conditions has only effected
the point $r=D$. In the context of the local chiral rotation it is
equally important to consider the wave--functions at $r=0$. As already
mentioned the solutions to the Dirac equation (\ref{h0}) are given
by spherical Bessel functions in the free case, $\Theta=0$. Except
of $j_0$ these vanish at the origin. In the case $\Theta\ne0$ we
adopt the boundary conditions $\Theta(0)=-n\pi$ thus no singularity
appears in the Dirac Hamiltonian (\ref{h0}) at $r=0$. Therefore
the radial parts of the quark wave--functions may be expressed as
linear combinations of the solutions to the free Dirac Hamiltonian.
{\it E.g.}
\be
g_\mu^{(G,+;1)}(r)=\sum_k V_{\mu k}[\Theta]N_k\sqrt{1+m/E_{kG}}\
j_G(k_{kG}r), \nonumber \\
f_\mu^{(G,+;1)}(r)=\sum_k V_{\mu k}[\Theta]N_k{\rm sign}(E_{kG})
\sqrt{1-m/E_{kG}}\ j_{G+1}(k_{kG}r)
\label{bescom}
\ee
where the eigenvectors $V_{\mu k}[\Theta]$ are obtained by
diagonalizing the Dirac Hamiltonian in the free basis. It should be
stressed that the use of the free basis is only applicable because
the point singularity hidden in ${\mbox{\boldmath $\tau$}} \cdot
{\bf \hat r}$ has disappeared. If singularities show up for
certain field configurations the basis for diagonalizing ${\cal H}$
has to be altered. This will be the central issue of the next
section.

\begin{table}
\tcaption{Properties of the two boundary conditions (\ref{bc1})
and (\ref{bc2}) in the baryon number zero sector. $f(r)$
represents an arbitrary radial function.}
{}~
\vskip0.5cm
\begin{tabular}{lcc}
Quantity& Condition (\ref{bc1})
& Condition  (\ref{bc2}) \\
\hline
$\sum_\mu\Psi_\mu^\dagger\beta\gamma_5
i{\mbox{\boldmath $\tau$}} \cdot{\bf \hat r} \Psi_\mu
{\rm erfc}\left(\left|\frac{\epsilon_\mu}{\Lambda}\right|\right)
{\rm sign}(\epsilon_\mu)=0$
& yes  & yes  \\
\hline
$\sum_\mu\Psi_\mu^\dagger{\mbox{\boldmath $\alpha$}} \cdot
\left({\mbox{\boldmath $\tau$}}\times{\bf \hat r}\right) \Psi_\mu
{\rm erfc}\left(\left|\frac{\epsilon_\mu}{\Lambda}\right|\right)
{\rm sign}(\epsilon_\mu)=0$
& no & yes  \\
\hline
$\sum_\mu\Psi_\mu^\dagger{\mbox{\boldmath $\alpha$}} \cdot
{\mbox{\boldmath $\tau$}} \Psi_\mu
{\rm erfc}\left(\left|\frac{\epsilon_\mu}{\Lambda}\right|\right)
{\rm sign}(\epsilon_\mu)=0$
& yes  & yes  \\
\hline
$\sum_\mu\Psi_\mu^\dagger{\mbox{\boldmath $\alpha$}} \cdot
{\bf \hat r} {\mbox{\boldmath $\tau$}}\cdot {\bf \hat r} \Psi_\mu
{\rm erfc}\left(\left|\frac{\epsilon_\mu}{\Lambda}\right|\right)
{\rm sign}(\epsilon_\mu)=0$
& yes  & yes  \\
\hline
$\tr \left(\beta f(r)\right)=0 $ & yes  & no \\
\hline
$\tr \left({\mbox{\boldmath $\gamma$}} \cdot
{\mbox{\boldmath $\tau$}} f(r)\right)=0 $ & yes  & yes  \\
\hline
$\langle\mu|\tau_i|\nu\rangle=0$ for $k_\mu\ne k_\nu$
& no & yes  \\
\end{tabular}
\end{table}

\section{The chirally rotated fermion determinant}

In the previous section we have demonstrated that the normalizable
solutions to the free Dirac equation with spherical boundary conditions
represent a pertinent basis for the diagonalization of the Dirac
Hamiltonian with the soliton present. This property is based on the
fact that the Hamiltonian (\ref{h0}) is free of singularities. In this
section we will explain how singularities appearing in a Dirac
Hamiltonian influence the choice of basis states. Let us for this
purpose consider the Hamiltonian for the chirally rotated
quark fields $\tilde \Psi=\Omega(\Theta)\Psi$
({\it cf.} eqns (\ref{wuyang})--(\ref{atheta}) and ref.\cite{re89a}):
\be
{\cal H}_R= \Omega(\Theta){\cal H}\Omega^\dagger(\Theta)&=&
\mbox{\boldmath $\alpha$}\cdot{\bf p}
+\beta m -\frac{1}{2}(\mbox{\boldmath $\sigma$}\cdot{\bf \hat{r}})
(\mbox{\boldmath $\tau$}\cdot{\bf \hat{r}})
\left(\Theta^\prime(r)-\frac{1}{r}\sin\Theta(r)\right) \nonumber \\
 & &-\frac{1}{2r}(\mbox{\boldmath $\sigma$}\cdot\mbox{\boldmath $\tau$})
\sin\Theta(r)-\frac{1}{r}\mbox{\boldmath $\alpha$}\cdot
({\bf \hat{r}}\times\mbox{\boldmath $\tau$})
\sin^2\left(\frac{\Theta(r)}{2}\right)
\label{hrot}
\ee
since in unitary gauge
$\Omega(\Theta)={\rm cos}(\Theta/2)
+i\gamma_5\mbox{\boldmath $\tau$}\cdot{\bf \hat{r}}\
{\rm sin}(\Theta/2)$.
Obviously the $\Theta$--dependence in the Hamiltonian has been
transferred to induced (axial--) vector meson fields. As expected the
rotated Hamiltonian, ${\cal H}_R$, contains an explicit singularity in
the $1/r\mbox{\boldmath $\alpha$}\cdot
({\bf \hat{r}}\times\mbox{\boldmath $\tau$})
\sin^2\left(\frac{\Theta(r)}{2}\right)$ term at $r=0$.
Additionally there are ``coordinate singularities" in the
expressions involving ${\bf \hat{r}}$. All these singularities appear
because the ``coordinate singularity" in $\Omega(\Theta)$ is
not compensated by corresponding values of the chiral angle $\Theta$.
Stated otherwise: the transformation $\Omega(\Theta)$ with
$\Theta(0)-\Theta(\infty)=-n\pi$ is topologically distinct from the
unit transformation. Although $\Omega(\Theta)$ represents a unitary
transformation it is then not astonishing that a numerical
diagonalization
\be
{\cal H}_R \tilde \Psi_\mu =\tilde\epsilon_\mu \Psi_\mu
\label{dhrot}
\ee
in the basis of the free Hamiltonian does ${\underline {\rm not}}$ render
the eigenvalues of the unrotated Hamiltonian, ${\cal H}$, ({\it i.e.}
$\tilde\epsilon_\mu\ne \epsilon_\mu)$ despite the relevant
matrix elements being finite.
This finiteness is merely due to the $r^2$ factor in the volume element.
One might suspect that the Hamiltonian ${\cal H}_{2R}=
\Omega(2\Theta){\cal H} \Omega^\dagger(2\Theta)$ obtained by a
$2\Theta$ rotation has the same spectrum as ${\cal H}$, since
${\cal H}_{2R}$ is free of singularities. Although this behavior is
exhibited by the numerical solution for the low--lying energy
eigenvalues, the topological character of the transformation has
drastic consequences for the states at the lower and upper ends of the
spectrum in momentum space. Adopting the same basis states for
diagonalizing ${\cal H}$ and ${\cal H}_{2R}$  one observes that the
eigenvalues of ${\cal H}_{2R}$ are shifted against those of ${\cal H}$,
{\it i.e.} the most negative energy eigenvalue is missing while an
additional one has popped up at the upper end of the spectrum. Up
to numerical uncertainties the eigenvalues in the intermediate region
agree for both ${\cal H}$ and ${\cal H}_{2R}$. This behavior is
sketched in figure 1 and repeats itself for
${\cal H}_{4R}=\Omega(4\Theta){\cal H}\Omega^\dagger(4\Theta)$.
Thus the chiral rotation represents another example of the
so--called ``infinite hotel story"\cite{ni83} which is an
interesting feature reflecting the topological character of this
transformation.

\begin{figure}
\begin{center}
\setlength{\unitlength}{1mm}
\begin{picture}(125,91)
\put(0,50){\line(1,0){105}}
\thicklines
\put(0,60){\line(1,0){15}}
\put(0,70){\line(1,0){15}}
\put(0,30){\line(1,0){15}}
\put(0,8){\line(1,0){15}}
\put(4,0){${\cal H}_{-2R}$}
\put(34,0){${\cal H}$}
\put(64,0){${\cal H}_{2R}$}
\put(94,0){${\cal H}_{4R}$}
\put(0,10){\line(1,0){15}}
\put(0,12){\line(1,0){15}}
\put(0,14){\line(1,0){15}}
\put(0,16){\line(1,0){15}}
\put(7.5,20){\circle*{0.8}}
\put(7.5,22){\circle*{0.8}}
\put(7.5,24){\circle*{0.8}}
\put(7.5,26){\circle*{0.8}}
\put(7.5,18){\circle*{0.8}}
\put(0,28){\line(1,0){15}}
\put(7.5,72){\circle*{0.8}}
\put(7.5,74){\circle*{0.8}}
\put(7.5,76){\circle*{0.8}}
\put(7.5,78){\circle*{0.8}}
\put(7.5,80){\circle*{0.8}}
\put(0,82){\line(1,0){15}}
\put(0,84){\line(1,0){15}}
\put(0,86){\line(1,0){15}}
\put(110,49){0}
\put(30,60){\line(1,0){15}}
\put(30,70){\line(1,0){15}}
\put(30,30){\line(1,0){15}}
\put(30,10){\line(1,0){15}}
\put(30,12){\line(1,0){15}}
\put(30,14){\line(1,0){15}}
\put(30,16){\line(1,0){15}}
\put(37.5,20){\circle*{0.8}}
\put(37.5,22){\circle*{0.8}}
\put(37.5,24){\circle*{0.8}}
\put(37.5,26){\circle*{0.8}}
\put(37.5,18){\circle*{0.8}}
\put(30,28){\line(1,0){15}}
\put(37.5,72){\circle*{0.8}}
\put(37.5,74){\circle*{0.8}}
\put(37.5,76){\circle*{0.8}}
\put(37.5,78){\circle*{0.8}}
\put(37.5,80){\circle*{0.8}}
\put(30,82){\line(1,0){15}}
\put(30,84){\line(1,0){15}}
\put(30,86){\line(1,0){15}}
\put(30,88){\line(1,0){15}}
\put(30,60){\line(1,0){15}}
\put(60,70){\line(1,0){15}}
\put(60,30){\line(1,0){15}}
\put(60,12){\line(1,0){15}}
\put(60,14){\line(1,0){15}}
\put(60,16){\line(1,0){15}}
\put(67.5,20){\circle*{0.8}}
\put(67.5,22){\circle*{0.8}}
\put(67.5,24){\circle*{0.8}}
\put(67.5,26){\circle*{0.8}}
\put(67.5,18){\circle*{0.8}}
\put(60,28){\line(1,0){15}}
\put(67.5,72){\circle*{0.8}}
\put(67.5,74){\circle*{0.8}}
\put(67.5,76){\circle*{0.8}}
\put(67.5,78){\circle*{0.8}}
\put(67.5,80){\circle*{0.8}}
\put(60,82){\line(1,0){15}}
\put(60,84){\line(1,0){15}}
\put(60,86){\line(1,0){15}}
\put(60,88){\line(1,0){15}}
\put(60,90){\line(1,0){15}}
\put(60,60){\line(1,0){15}}
\put(90,60){\line(1,0){15}}
\put(110,59){$\epsilon_{\rm val}$}
\put(90,70){\line(1,0){15}}
\put(110,69){$+m$}
\put(110,29){$-m$}
\put(90,30){\line(1,0){15}}
\put(90,14){\line(1,0){15}}
\put(90,16){\line(1,0){15}}
\put(97.5,20){\circle*{0.8}}
\put(97.5,22){\circle*{0.8}}
\put(97.5,24){\circle*{0.8}}
\put(97.5,26){\circle*{0.8}}
\put(97.5,18){\circle*{0.8}}
\put(90,28){\line(1,0){15}}
\put(97.5,72){\circle*{0.8}}
\put(97.5,74){\circle*{0.8}}
\put(97.5,76){\circle*{0.8}}
\put(97.5,78){\circle*{0.8}}
\put(97.5,80){\circle*{0.8}}
\put(90,82){\line(1,0){15}}
\put(90,84){\line(1,0){15}}
\put(90,86){\line(1,0){15}}
\put(90,88){\line(1,0){15}}
\put(90,90){\line(1,0){15}}
\put(90,92){\line(1,0){15}}
\end{picture}
\end{center}
\noindent
\fcaption{A schematic plot of the spectrum of the rotated
Hamiltonian ${\cal H}_{nR}=\Omega(n\Theta){\cal H}
\Omega^\dagger(n\Theta)$.}
\end{figure}

Let us now return to the problem of diagonalizing ${\cal H}_R$ and restrict
ourselves for the moment to the channel $G^\Pi=0^+$. At $r=0$
the chiral rotation
\be
\Omega(r=0)=-i(\mbox{\boldmath $\tau$}\cdot{\bf \hat{r}})\gamma_5
\label{globalrot}
\ee
obviously exchanges upper and lower components of Dirac
spinors\footnote{In the standard representation
$\gamma_5=\pmatrix{0 & 1\cr 1 &0\cr}$.}. The corresponding wave--functions
are given by
\be
\Omega(r=0)\Psi_\mu^{(0,+)}(r=0)=
\pmatrix{-if_\mu^{(0,+;1)}(r=0)|0\frac{1}{2}00\rangle \cr
g_\mu^{(0,+;1)}(r=0) |1\frac{1}{2}00\rangle \cr}
\label{psi0rot}
\ee
which obviously cannot be represented by the free basis since
$g_\mu^{(0,+;1)}(r=0)\ne 0$, in general. However, the lower components
of the eigenstates of the free Hamiltonian in the $G^\Pi=0^+$ channel
have this property  ({\it cf.} eqns (\ref{psipos},\ref{freesol})). We
therefore apply the rotation (\ref{globalrot}) to the eigenstates of
the free Hamiltonian as well, resulting in a new basis given by
\be
N_k\pmatrix {-i\ {\rm sign}(E)\sqrt{1-E/m}\
j_1(kr)|0\frac{1}{2}00\rangle \cr
\sqrt{1+E/m}\ j_0(kr)|1\frac{1}{2}00\rangle \cr}.
\label{basis0p}
\ee
This is, of course, no longer a solution to the free Dirac equation.
Moreover, a single component of this spinor does not even solve the
free Klein Gordan equation. At $r=D$ the chiral rotation equals unity.
Thus we demand the discretization condition $j_1(kD)=0$ according
to eqn (\ref{bc1}) {\it i.e.} the momenta are taken from the set
$\{k_{n1}\}$. With this basis we have succeeded in eliminating
the singularities at $r=0$ and keeping track of the boundary conditions
at $r=D$. We are thus enabled to numerically diagonalize ${\cal H}_R$.
When comparing with the eigenvalues of ${\cal H}$ we again encounter a
form of the ``infinite hotel story": one state is missing in the negative
part of the spectrum while an additional shows up in the positive part.
The missing state turns out to be the one at the upper end of the
negative Dirac sea, {\it i.e.} $E\approx-m$. Then it is important to
note that in addition to the states with finite $k$ the basis
(\ref{basis0p}) together with the boundary condition $j_1(kD)=0$ also
allows for the ``constant state" with $k=0$. In the continuum limit
($D\rightarrow\infty$) this state is absent. Including, however, this
state for finite $D$ in the process of diagonalizing ${\cal H}_R$
finally renders the missing state. This is not surprising since
application of the inverse chiral rotation $\Omega^\dagger(r=0)$ onto
this ``constant state" leads to an eigenstate of the free Dirac
Hamiltonian with the eigenvalue $-m$. It should be remarked that for
the free unrotated problem a ``constant state" with eigenvalue
$-m$ is only allowed in the $G^\Pi=1^-$ channel. Although
$\Omega(\Theta)$ does commute with the grand spin operator its
topological character mixes various grand spin channels via the
boundary conditions.

Accordingly the diagonalization of ${\cal H}_R$ in the $G^\Pi=0^-$ channel
demands the basis states
\be
N_k\pmatrix {i\ {\rm sign}(E)\sqrt{1-E/m}\
j_0(kr)|1\frac{1}{2}00\rangle \cr
\sqrt{1+E/m}\ j_1(kr)|0\frac{1}{2}00\rangle \cr}
\label{basis0m}
\ee
with the boundary condition $j_1(kD)=0$ in order to be compatible with
the Kahana--Ripka\cite{ka84} diagonalization of ${\cal H}$. The
additional ``constant state" needed here corresponds to a state with
eigenvalue $+m$ of the free unrotated Hamiltonian\footnote{The additional
``constant states" thus do not alter the trace of the Hamiltonian.}.

We have finally been able to diagonalize the chirally rotated
Hamiltonian ${\cal H}_R$ in the $G=0$ sectors by very tricky means. It
should also be kept in mind that there are now additional states at the
upper (from $0^+$) and lower (from $0^-$) ends of the spectrum which
do not possess ``counterstates" in the $G=0$ part of the spectrum
in the unrotated problem. For the dynamics of the problem they are of
no importance because their contribution to physical quantities is
damped by the regularization. However, their existence reflects the
topological character of the chiral rotation.

In the other channels {\it i.e.} $G\ge1$ we have unfortunately not
been able to construct a set of basis states which rendered the
eigenvalues of the unrotated Hamiltonian along the approach
described above. In the $G=0$ sector we have already seen that a
``global rotation" $-i(\mbox{\boldmath $\tau$}\cdot{\bf
\hat{r}})\gamma_5$ is needed for the basis states in order to
accommodate the boundary conditions at $r=0$. Furthermore a mixture
of different grand spin channels appears since this ``global rotation"
deviates from unity at $r=D$. This problem can be avoided by
defining a basis in the topologically non--trivial sector via
\be
\tilde \Psi_{\mu 0}=\Omega^\dagger(\phi)\Psi_{\mu 0}
\label{topbasis}
\ee
with $\Psi_{\mu 0}$ being the solutions to the free unrotated
Hamiltonian. $\phi$ represents an auxiliary  radial function satisfying
the boundary conditions $\phi(0)=-\pi$ and $\phi(D)=0$. {\it E.g.}
we may take
\be
\phi(r)=-\pi\left(1-\frac{r}{D}\right)
{\rm exp}\left(-tmr\right)
\label{ansatzphi}
\ee
with $t$ being a free parameter. The diagonalization of ${\cal H}_R$
in the basis $\tilde \Psi_0$ is equivalent to diagonalizing
\be
&&\hspace{2cm}
\mbox{\boldmath $\alpha$}\cdot{\bf p}
+m \beta \left(\cos\phi(r)+i\gamma_5
(\mbox{\boldmath $\tau$}\cdot{\bf \hat{r}})
\sin\phi(r)\right) \nonumber \\
&&\hspace{1cm}
+\frac{1}{2}\left[\phi^\prime(r)-\Theta^\prime(r)
+\frac{1}{r}\sin(\Theta(r)-\phi(r))\right]
(\mbox{\boldmath $\sigma$}\cdot{\bf \hat{r}})
(\mbox{\boldmath $\tau$}\cdot{\bf \hat{r}}) \nonumber \\
&&+\frac{1}{2r}\sin(\phi(r)-\Theta(r))
(\mbox{\boldmath $\sigma$}\cdot\mbox{\boldmath $\tau$})
-\frac{1}{2r}\left[1-\cos(\Theta(r)-\phi(r))\right]
\mbox{\boldmath $\alpha$}\cdot
({\bf \hat{r}}\times\mbox{\boldmath $\tau$})
\label{hphi}
\ee
in the standard basis $\{\Psi_{\mu0}\}$. At this point
it should be stressed again that $\phi$ is not a dynamical field
but merely an auxiliary field which transforms the basis such as
to eliminate the singularities from the Dirac Hamiltonian. This
property is completely determined by the boundary values of
$\phi$. Thus the results ought to be independent of the parameter
$t$.

For the numerical investigation we have first considered the energy
eigenvalues of ${\cal H}_R$ employing the rotated basis (\ref{topbasis})
for the diagonalization and compared to the eigenvalues of ${\cal H}$.
In a wide range $t\approx0.1\ldots1.4$ the difference for the eigenvalues
which are contained in the interval $(-5m,5m)$ is negligible, {\it i.e.}
less than tenth of a percent. For the higher grand spin channels the
window for $t$ which yields reasonable agreement lies somewhat higher
$t\approx0.8\ldots1.6$. Summing up the energy eigenvalues according to
eqn (\ref{evac}) in order to obtain the vacuum part of the soliton energy
the deviation compared to the unrotated formulation is about $0.5$MeV.
This is, of course, negligibly small since the inherited mass scale is
given by the constituent quark mass, $m$ which is of the order of several
hundred MeV. By using the locally transformed basis (\ref{topbasis})
the wave--functions corresponding to the eigenstates of ${\cal H}_R$
agree reasonably well with the rotated wave--functions $\Omega(\Theta)
\Psi_\mu$ of the original Dirac Hamiltonian ${\cal H}$. It should be
noted that a large number of momentum states is required to numerically
gain this result. This is not surprising since in order to represent
the functional unity an infinite number of momentum states is needed.

Before turning to the detailed discussion of the equation of motion in
the rotated system we would like to mention that the self--consistent
profile being obtained from the unrotated problem also minimizes the
soliton mass in the chirally rotated frame. Stated otherwise: each
change in this profile function leads to an increase of the energy
obtained from the eigenvalues of ${\cal H}_R$.

As in the formulation with in the unrotated frame the equation
of motion is gained by extremizing the energy functional (\ref{etot}).
The energy eigenvalues in the rotated frame, however, exhibit
a different functional dependence on the chiral field. The functional
derivative of these eigenvalues with respect to the chiral angle
reads
\be
\frac{\delta\epsilon_\mu}{\delta\Theta(r)}&=&
\int d\Omega\ \Big\{\frac{1}{2}\Big(\frac{\partial}{\partial r}
+\frac{1}{r}{\rm cos}\Theta(r)\Big) r^2 \tilde\Psi_\mu^\dagger({\bf r})
(\mbox{\boldmath $\sigma$}\cdot{\bf \hat{r}})
(\mbox{\boldmath $\tau$}\cdot{\bf \hat{r}})\tilde\Psi_\mu({\bf r})
\nonumber \\ &&  \qquad
-\frac{r}{2}{\rm cos}\Theta(r)\tilde\Psi_\mu^\dagger({\bf r})
\mbox{\boldmath $\sigma$}\cdot\mbox{\boldmath $\tau$}
\tilde\Psi_\mu({\bf r})
-\frac{r}{2}{\rm sin}\Theta(r)\tilde\Psi_\mu^\dagger({\bf r})
\mbox{\boldmath $\alpha$}\cdot
({\bf \hat{r}}\times\mbox{\boldmath $\tau$})
\tilde\Psi_\mu({\bf r})\Big\}
\label{diffrot}
\ee
with $\tilde\Psi_\mu({\bf r})$ being the eigenstates of the rotated
Dirac Hamiltonian (\ref{hrot}). The derivatives (\ref{diffrot})
enter the stationary condition for the energy functional resulting
in the equation of motion
\be
A_L(r)+A_T(r){\rm cos}\Theta(r)-V(r){\rm sin}\Theta(r)=0.
\label{eqmrot}
\ee
In order to display the radial functions $A_L,A_T$ and $V$ it
is convenient to introduce the charge density $\tilde\rho_C=
\langle q({\bf x}) q({\bf y})^\dagger\rangle=
\tilde\rho_C^{\rm val}+\tilde\rho_C^{\rm vac}$ involving the
eigenstates $\tilde\Psi_\mu$ of ${\cal H}_R$ ({\it cf.} eqn
(\ref{density}))\cite{re89}:
\be
\tilde\rho_C^{\rm val}({\bf x},{\bf y}) & = & \sum _\mu
\tilde\Psi_\mu({\bf x})\eta_\mu \tilde\Psi_\mu^\dagger({\bf y})
{\rm sign} (\epsilon_\mu)
\nonumber \\*
\tilde\rho_C^{\rm vac}({\bf x},{\bf y}) & = &
\frac{-1}{2}\sum_\mu \tilde\Psi_\mu({\bf x})
{\rm erfc}\left(\left|\frac{\epsilon_\mu}{\Lambda}\right|\right)
\tilde\Psi_\mu^\dagger({\bf y}) {\rm sign} (\epsilon_\mu) .
\label{densrot}
\ee
The radial functions $A_L,A_T$ and $V$ are of axial--($A_{L,T}$)
and vector($V$) character
\be
A_L(r)&=&\frac{1}{r}\frac{\partial}{\partial r}\
{\rm tr}\ \int d\Omega\ r^2\ \tilde\rho_C({\bf r},{\bf r})
(\mbox{\boldmath $\sigma$}\cdot{\bf \hat{r}})
(\mbox{\boldmath $\tau$}\cdot{\bf \hat{r}})
\label{defal} \\
A_T(r)&=&{\rm tr}\ \int d\Omega\ \tilde\rho_C({\bf r},{\bf r})
\left[(\mbox{\boldmath $\sigma$}\cdot{\bf \hat{r}})
(\mbox{\boldmath $\tau$}\cdot{\bf \hat{r}})-
(\mbox{\boldmath $\sigma$}\cdot\mbox{\boldmath $\tau$})\right]
\label{defat} \\
V(r)&=&{\rm tr}\ \int d\Omega\ \tilde\rho_C({\bf r},{\bf r})
\mbox{\boldmath $\alpha$}\cdot
({\bf \hat{r}}\times\mbox{\boldmath $\tau$}).
\label{defv}
\ee
It should be noted that $A_L$ does not depend on the auxiliary field
$\phi$. It is then straightforward to verify that for any meson
configuration $A_L$ and $A_T$ satisfy the relation
\be
A_L(r=0)=(-1)^kA_T(r=0)
\label{aorigin}
\ee
where $k$ is defined by the value of the auxiliary field $\phi$ at
the origin $\phi(r=0)=k\pi$. The vector type radial function
$V$ vanishes at the origin. Thus the equation of motion
(\ref{eqmrot}) together with the relation (\ref{aorigin}) yield
the boundary condition $\Theta(r=0)=(2n+1)\pi$ for $k=1$. This is
stronger than the boundary condition derived from the original equation
of motion (\ref{eqm}) which also allows for even multiples of $\pi$
for $\Theta(r=0)$ since $\tr \int d\Omega\ \rho_S({\bf r}=0,
{\bf r}=0)i\gamma_5 \mbox{\boldmath $\tau$}\cdot{\bf \hat{r}}=0$.
Assuming the Kahana--Ripka\cite{ka84} boundary conditions for the
unrotated basis states $\Psi_{\mu0}$ similar considerations for $r=D$
show that $\Theta(r=D)=2l\pi$ since $A_L(r=D)=-A_T(r=D)$ as long as
$\phi(D)=0$. Obviously the topological charge associated with the
chiral field in the hedgehog {\it ansatz} $(\Theta(r=0)-\Theta(r=D))/\pi$
can assume odd values only when $k$ is odd. Especially $\phi\equiv0$
is prohibited in the case of unit baryon number. Thus the study of the
boundary conditions in the chirally transformed system corroborates the
conclusion drawn from investigating the eigenvalues and --states
of ${\cal H}_R$ that it is mandatory to also transform the basis
spinors and in particular the boundary conditions to the topological
non--trivial sector.

Before turning to the discussion of the numerical treatment
of eqn (\ref{eqmrot}) we would like to make the remark
that substituting the transformation $\Psi_\mu({\bf r})=\Omega^\dagger
(\Theta)\tilde\Psi_\mu({\bf r})$ into the original equation of
motion (\ref{eqm}) does not result in the relation (\ref{diffrot})
but rather yields the constraint
\be
0=\sum_\mu\left(\eta_\mu {\rm sign}(\epsilon_\mu)
-\frac{1}{2}{\rm erfc}\left(\left|
\frac{\epsilon_\mu}{\Lambda}\right|\right)\right)
\bar{\tilde\Psi}_\mu({\bf r}) \gamma_5
\mbox{\boldmath $\tau$}\cdot{\bf \hat{r}}
\tilde\Psi_\mu({\bf r}).
\label{constr}
\ee
Thus the equation of motion (\ref{eqmrot}) cannot be obtained by
transforming the states $\Psi_\mu$ but only by employing the
Dirac equation in the rotated frame (\ref{dhrot}) to extremize the
energy functional. The constraint (\ref{constr}) does not represent
an over--determination of the the system since infinitely many
states are involved.

Due the transcendent character of the equation of motion in the rotated
frame (\ref{eqmrot}) solutions cannot be obtained for arbitrary values
of the radial functions $A_L,A_T$ and $V$. {\it E.g.} for large
distances $\Theta\rightarrow0$ requires $|A_L|\ge|A_T|$ in order
to find a solution to eqn (\ref{eqmrot}). Thus the treatment of the
NJL soliton in the chirally rotated frame is not very well suited for
an iterative procedure to find the self--consistent solution. The reason
is that a small deviation of the radial functions $A_L,A_T$ and $V$ from
those corresponding to this solution can render eqn (\ref{eqmrot})
indissoluble for $\Theta(r)$. Then it is not unexpected that at
large distances the solution to the rotated equation of motion
(\ref{eqmrot}) becomes unstable and the original profile function
cannot be reproduced for $r\ge2$fm. For smaller values of $r$ the
original profile function is well reproduced. In figure 2 this behavior
is displayed. The self--consistent solution to eqn (\ref{eqm})
serves as ingredient to evaluate the radial functions $A_L,A_T$ and $V$.
The solution to eqn (\ref{eqmrot}) is then constructed and compared
to the original profile function.
\centerline{\hskip -1.5cm
\psfig{figure=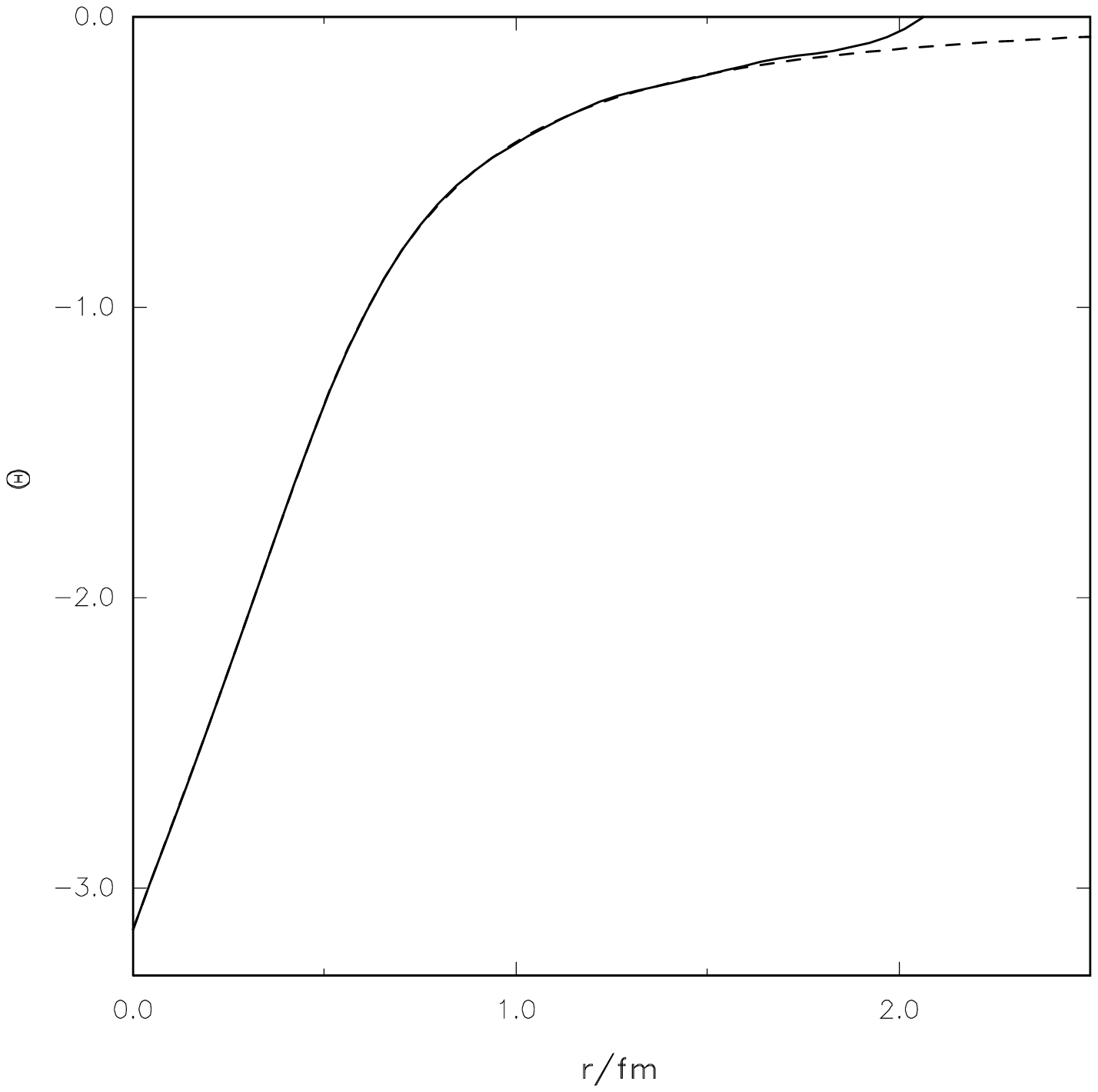,height=10.0cm,width=16.0cm}}
\fcaption{Comparison of the self--consistent profile in the unrotated
formulation (dashed line) and the solution to eqn (\ref{eqmrot})
(solid line).}

\section{Conclusions}

We have investigated the role of chiral transformations for the
evaluation of fermion determinants. When these transformations
are topologically trivial they provide a useful tool to evaluate
the chiral anomaly. Furthermore they can be used to demonstrate the
equivalence between the hidden gauge and massive Yang--Mills approaches
to vector mesons. A generalization to the case when chiral fields
have a topological charge different from zero is not straightforward.
Even though the special transformation we have been considering is
unitary its topological character prevents the eigenvalues and
--vectors of the original Dirac Hamiltonian to be regained from
the rotated Hamiltonian unless the boundary conditions are transformed
to the topologically non--trivial sector accordingly.
Furthermore we have observed that the stationary conditions to
the static energy functional in the topologically distinct sectors
are ${\underline {\rm not}}$ related by the transformation of
the equation of motion. The boundary conditions for the chiral field
obtained from the stationary condition have been found being invariant
under the chiral rotation only when the basis quark fields are taken
from the topological sector associated with the chiral transformation.
Diagonalization of the rotated Dirac Hamiltonian in this basis can
be reformulated into a problem where the induced vector fields
(\ref{indvec}) belong to the topologically trivial sector
({\it cf.} eqn (\ref{hphi})). In order to diagonalize the resulting
operator (\ref{hphi}) the standard basis\cite{ka84,we92} may be
employed.

These explorations on chiral transformation properties of the
fermion determinant when the Dirac Hamiltonian contains
topologically non--trivial (axial--) vector mesons may prove
to be very helpful when considering the chiral invariant
elimination of the axial--vector fields as described at the end
of section 2\cite{kay84a}. In that case the chirally rotated Dirac
Hamiltonian becomes as simple as
\be
{\cal H}_R=\mbox{\boldmath $\alpha$}\cdot{\bf p} +\beta m
+\frac{G(r)}{r}\mbox{\boldmath $\alpha$}\cdot
({\bf \hat{r}}\times\mbox{\boldmath $\tau$}).
\label{hrvec}
\ee
Here $G(r)$ refers to a dynamical field and should not be confused
with the induced vector field discussed in eqn (\ref{gtheta}).
The chiral field then only appears in the purely mesonic part
of the energy functional
\be
E_{\rm m}=\frac{2\pi}{G_2}\int dr \left\{
\left(G(r)+1-{\rm cos}\Theta(r)\right)^2+\frac{1}{2}r^2
\Theta^\prime(r)+{\rm sin}^2\Theta(r)\right\}.
\label{emvec}
\ee
$\Theta(r=0)=-\pi$ then implies that $G(r=0)=-2$\cite{ja88,me88}.
One thus has to deal with a Dirac Hamiltonian which contains
a topologically non--trivial vector meson field. Thus (\ref{hrvec})
cannot be treated using the standard basis\cite{ka84,we92} but
rather by employing techniques which are analogous to those developed
in section 4. Investigations in this direction are in progress.

Finally we would also like to remark that this kind of singularities
does not only appear when the Dirac Hamiltonian is considered. Such
topological defects have also caused problems when fluctuations off
vector meson solitons were investigated\cite{sch89}. In that case the
boundary conditions for the vector meson fluctuations had to undergo
a special gauge transformation which corresponds to the transformation
of the basis quark spinors described here (\ref{topbasis}).

\vskip2cm

\nonumsection{Acknowledgement}
We would like to thank U. Z\"uckert for helpful
contributions in the early stages of this work.

\end{document}